\documentclass[oldversion]{aa}
\usepackage{graphicx}
\usepackage{natbib}
\usepackage{hyperref}
\begin{document}

   \title{High-contrast Imaging with \textit{Spitzer}: Deep Observations of Vega, Fomalhaut, and $\epsilon$ Eridani
%\thanks{Based on observations collected at the European Southern Observatory, 
%Chile (NACO SDI commissioning run, February 2004).}
}

  % \subtitle{High-contrast imaging with Spitzer}

   \author{Markus Janson\inst{1} \and
          Sascha P. Quanz\inst{2} \and
          Joseph C. Carson\inst{3} \and
          Christian Thalmann\inst{2} \and
          David Lafreni{\`e}re\inst{4} \and
          Adam Amara\inst{2}
	  }

   \offprints{Markus Janson}

   \institute{Department of Astronomy, Stockholm University, Stockholm, Sweden\\
              \email{markus.janson@astro.su.se}
         \and
             Institute for Astronomy, ETH Zurich, Zurich, Switzerland\\
	\email{sascha.quanz@astro.phys.ethz.ch, thalmann@phys.ethz.ch, adam.amara@phys.ethz.ch}
	\and
	    Department of Physics and Astronomy, College of Charleston, Charleston, SC, USA\\
	\email{carsonjc@cofc.edu}
	\and
	    Department of Physics, University of Montreal, Montreal, QC, Canada\\
	\email{david@astro.umontreal.ca}
             }

   \date{Received ---; accepted ---}

   \abstract{Stars with debris disks are intriguing targets for direct imaging exoplanet searches, both due to previous detections of wide planets in debris disk systems, as well as commonly existing morphological features in the disks themselves that may be indicative of a planetary influence. Here we present observations of three of the most nearby young stars, that are also known to host massive debris disks: Vega, Fomalhaut, and $\epsilon$~Eri. The \textit{Spitzer} Space Telescope is used at a range of orientation angles for each star, in order to supply a deep contrast through angular differential imaging combined with high-contrast algorithms. The observations provide the opportunity to probe substantially colder bound planets (120--330~K) than is possible with any other technique or instrument. For Vega, some apparently very red candidate point sources detected in the 4.5~$\mu$m image remain to be tested for common proper motion. The images are sensitive to $\sim$2~$M_{\rm jup}$ companions at 150~AU in this system. The observations presented here represent the first search for planets around Vega using \textit{Spitzer}. The upper 4.5~$\mu$m flux limit on Fomalhaut b could be further constrained relative to previous data. In the case of $\epsilon$~Eri, planets below both the effective temperature and the mass of Jupiter could be probed from 80~AU and outwards, although no such planets were found. The data sensitively probe the regions around the edges of the debris rings in the systems where planets can be expected to reside. These observations validate previous results showing that more than an order of magnitude improvement in performance in the contrast-limited regime can be acquired with respect to conventional methods by applying sophisticated high-contrast techniques to space-based telescopes, thanks to the high degree of PSF stability provided in this environment.}


\titlerunning{High-contrast Imaging with \textit{Spitzer}}
\authorrunning{M. Janson et al.}

   \maketitle
%
%________________________________________________________________

\section{Introduction}
\label{s:intro}

A large fraction of the extrasolar planets that have been directly imaged to date reside in systems with massive debris disks \citep[e.g.][]{marois2008,lagrange2010,rameau2013}. This may imply some correlation between at least wide massive planets and such disks. Furthermore, many debris disks show signs in both infrared excess \citep[e.g.][]{hillenbrand2008,trilling2008} and spatially resolved imaging \citep[e.g.][]{schneider1999,kalas2005} of having ring-like structures with inner gaps and cavities, sometimes with eccentric shapes. While alternative possibilities exist for explaining these structures \citep[e.g.][]{lyra2013}, such rings may be shaped by planets orbiting near the ring edges \citep[e.g.][]{quillen2006}. It is therefore interesting to study the regions close to the edges in particular detail \citep[e.g.][]{apai2008,janson2013a}. While many planetary imaging surveys are performed at JHK-band ($\sim$1--2.5~$\mu$m) wavelengths \citep[e.g.][]{chauvin2010,nielsen2013,brandt2014}, there is considerable interest in studying planets at longer wavelengths, in the LM-band range ($\sim$3--5~$\mu$m). For ground-based telescopes, one of the reasons for this is the enhanced Point Spread Function (PSF) stability \citep[e.g.][]{kasper2007,janson2008,heinze2008}, but a more general reason is the fact that only warm planets of $\sim$400~K effective temperatures and higher emit any significant flux at JHK-band. At lower temperatures, the flux in this region drops rapidly, while significant flux remains at longer wavelengths. Indeed, a wealth of results, including some of the planet detections mentioned above, have been achieved in the 3--5~$\mu$m wavelength regime. However, a dominant limiting factor, in particular for ground-based telescopes, is the high level of thermal background noise that occurs even for cooled instruments.

$\epsilon$~Eri is a K2V-type star at a distance of 3.2~pc, Fomalhaut is an A4V star at 7.7~pc, and Vega is an A0V star at 7.7~pc \citep{perryman1997,vanleeuwen2007}. They all have large debris disks with inner gaps, and they all have ages of a few to several hundred Myr, as discussed in subsequent sections. This makes them exceptional targets for planet imaging studies, and indeed, a large number of dedicated imaging studies of these targets have been performed as the field has developed \citep[e.g.][and the many others mentioned in the discussion of individual objects below]{macintosh2003,metchev2003,janson2007}. Furthermore, they have all had candidate planetary companions inferred around them. Fomalhaut~b is a visible-light point source observed with the Hubble Space Telescope (HST), as reported in \citet{kalas2008}. The point source corresponds to a real physical object bound to the system, but its exact nature remains unclear \citep[e.g.][and this paper]{janson2012,galicher2013,kalas2013}. $\epsilon$~Eri~b was reported as a radial velocity signature with a $\sim$7~yr period by \citet{hatzes2000}, whose semi amplitude suggested a mass in the jovian range. Later studies have supported this statement through astrometric measurements of the host star \citep{benedict2006,reffert2011}. However, subsequent radial velocity studies with better precision have been unable to verify the existence of the planet \citep[e.g.][]{anglada2012,zechmeister2014}, implying a spurious detection or significantly different orbital parameters than those originally reported. All systems have had general predictions of planets from the disk morphology, and in the case of Vega, a rather specific prediction on the precise location of a planet has been made on the basis of what was interpreted as resonant features in the disk \citep{wilner2002,deller2005}. The underlying image however was based on interferometry with relatively limited coverage of the UV plane, and subsequent studies have not verified this morphology \citep[e.g.][]{hughes2012}.

In pioneering work by \citet{marengo2006}, it was shown that \textit{Spitzer} could place stronger limits on wide separation planets in the $\epsilon$~Eri system than any other existing facility. The detection limits were further improved for this system in \citet{marengo2009}, where Fomalhaut was also studied to a similar degree of sensitivity. Subsequently, in \citet{janson2012} a dedicated high-contrast observational and data reduction scheme was applied which further substantially enhanced the detection limit of \textit{Spitzer} in the contrast-limited regime. This made it clear that the \textit{Spitzer} space telescope could be efficiently used for high-contrast imaging at 4.5~$\mu$m, although the small aperture size of the telescope (0.85~m) limits the angular separation from the central star down to which the telescope can efficiently probe. The three stars described above are all very nearby, which places even modest physical separations at large angular separations, and they are thus ideal targets to study with \textit{Spitzer}, allowing for much colder planets to be detected than are available with any other existing technique or telescope. Here we report on a dedicated survey for acquiring deep observations at multiple orientation angles of the telescope for Vega, Fomalhaut, and $\epsilon$~Eri, and the results attained from applying angular differential imaging and high-contrast algorithms to the data. In Sect. \ref{s:obs}, we will describe the observing strategy and the basic data reduction, and in Sect. \ref{s:psf} we will outline the PSF subtraction methods used in the study. The individual results for the three targets are described in section \ref{s:results}. We summarize the overall results and conclusions in Sect. \ref{s:summary}.

\section{Observations and Data Reduction}
\label{s:obs}

All observations in this work were acquired using the IRAC camera on the \textit{Spitzer} Space Telescope. The three targets $\epsilon$~Eri (03:32:55.84, -09:27:29.7)\footnote{All coordinates are in hh:mm:ss, dd:mm:ss and J2000.0 format.}, Fomalhaut (22:57:39.05, -29:37:20.1), and Vega (18:36:56.34, +38:47:01.3) were observed with an identical observation scheme, building on the procedure developed for Fomalhaut as described in \citet{janson2012}. Each target was observed on eight different occasions across \textit{Spitzer} cycle 9, which for regular programs ran from January through December of 2013. The eight occasions were spread out as much as possible within and across the observing windows available for each target during the cycle. This was done in order to optimize the observations for angular differential imaging (ADI) performance. ADI is a technique in which the target star is observed at several different rotation angles of the telescope, so that it can act effectively as a PSF reference for itself, as any off-axis sources will be located at different position angles in each data set. The wider the spread in rotation angles at which the system is observed, the smaller separations can be usefully probed with ADI. Since \textit{Spitzer} is unable to actively roll, we take advantage of the fact that it exhibits a nominal roll across a year, and so a maximally spread out scheduling of the various observations facilitates an optimal ADI performance. Each of the eight visits for a given target consists of a series of 96 consecutive exposures, of which 48 are in the 3.6~$\mu$m and 48 in the 4.5~$\mu$m channel. A 12-point Roleaux dither pattern is used for oversampling purposes. Individual exposures have 10.4~s of effective integration time, leading to a total frame time of 12~s. A log of the dates and position angles for the 24 target visits (eight visits each for three targets) is provided in Table \ref{t:log}. 

\begin{table}[htb]
\caption{Log of observing dates and angles}
\label{t:log}
\centering
\begin{tabular}{lrrr}
\hline
\hline
Target & AOR & Date & PA (deg) \\
\hline
$\epsilon$~Eri	& 47936512 & 2013-03-19 & 58.1 \\
$\epsilon$~Eri	& 47936768 & 2013-03-31 & 64.8 \\
$\epsilon$~Eri	& 47937024 & 2013-04-08 & 69.0 \\
$\epsilon$~Eri	& 47937280 & 2013-04-22 & 75.5 \\
$\epsilon$~Eri	& 47937536 & 2013-10-22 & -106.0 \\
$\epsilon$~Eri	& 47937792 & 2013-10-29 & -102.8 \\
$\epsilon$~Eri	& 47938048 & 2013-11-08 & -98.2 \\
$\epsilon$~Eri	& 47938304 & 2013-11-27 & -88.4 \\
Fomalhaut	& 47938560 & 2013-01-19 & 59.6 \\
Fomalhaut	& 47938816 & 2013-01-23 & 61.1 \\
Fomalhaut	& 47939072 & 2013-01-26 & 62.2 \\
Fomalhaut	& 47939328 & 2013-02-10 & 67.4 \\
Fomalhaut	& 47939584 & 2013-08-06 & -115.0 \\
Fomalhaut	& 47939840 & 2013-08-10 & -113.5 \\
Fomalhaut	& 47940096 & 2013-08-22 & -109.4 \\
Fomalhaut	& 47940352 & 2013-08-28 & -107.1 \\
Vega 	& 47934720 & 2013-06-11 & -92.2 \\
Vega 	& 48354816 & 2013-06-17 & -96.9 \\
Vega 	& 47934976 & 2013-07-04 & -111.8 \\
Vega 	& 47935232 & 2013-08-21 & -156.7 \\
Vega 	& 47935488 & 2013-09-08 & -175.6 \\
Vega 	& 47935744 & 2013-10-16 & 143.8 \\
Vega 	& 47936000 & 2013-11-27 & 104.7 \\
Vega 	& 47936256 & 2013-12-05 & 97.8 \\
\hline
\end{tabular}
%\tablenotetext{a}{No notes at the moment.}
\end{table}

For data reduction purposes, we used the post-BCD data from the \textit{Spitzer} Heritage Archive, for which the fundamental reduction steps such as flat fielding and dark subtraction have already been performed. Although some bad pixel identification and removal has also been performed on these data, there were a few residual bad pixels present in the frames, hence we ran an additional bad pixel removal scheme in the same way as was used and described in \citet{janson2012}, by identifying outliers from the median of each quadruple set of frames corresponding to one particular dither position in an observing sequence. Next, the frames from all runs for a given target were registered to a common center, which cannot be done on the PSF core as it is always saturated in the images. Instead, this was done by cross-correlating each frame with a 180$^{\rm o}$ rotated version of itself. For this purpose, only the six PSF spider arms were used, selecting 10 pixel wide strips from 20 to 70 pixels separation on each spider arm. The resulting centers were checked visually and also compared with an approximate center based on Gaussian centroiding on a strongly low-pass filtered version of each frame. In the vast majority of cases, the registering yielded a center that was accurate and consistent between the various checks, but in a few cases the result of the automatic centering was visibly off center; in these cases, we redid the automatic centering with a different cross-correlation window, typically from 10 to 30 pixel separation on the spider arms. Following this procedure, a satisfactory centering was found for the full set of images. Although the comparison strategy of Gaussian centering on a low-pass filtered version of the frame is considered less precise due to the broadening of the PSF in the process, the standard deviation in the difference between the coordinates found from the cross-correlation method versus the comparison method is only $\sim$120~mas. Thus, we can expect the cross-correlation method to have a precision for individual frames of at least this level or better, which corresponds to 7\% of the PSF Full Width at Half Maximum (FWHM; 1.72\arcsec\ at 4.5~$\mu$m). All frames were then translated to their common center and oversampled to a final pixel scale of 300 mas/pixel, using spline interpolation. 

To summarize the spatial dimensions of IRAC, the average FWHM given by the IRAC manual\footnote{http://irsa.ipac.caltech.edu/data/SPITZER/docs/irac/} is 1.66\arcsec\ at 3.6~$\mu$m and 1.72\arcsec\ at 4.5~$\mu$m. The field of view is approximately 5.2\arcmin\ on each side, and the pixel scale in our oversampled frames is 300~mas/pixel for both bands.

% Maybe note something about the fact that we tried our own BCD with identical results

\section{PSF subtraction}
\label{s:psf}

Each frame has an individual PSF reference constructed for it, using the library of frames taken of the same target at different position angles. The PSF reference construction for high-contrast purposes is typically done in one of two different ways in the literature (or variations thereof): One method is Locally Optimized Combination of Images \citep[LOCI, see][]{lafreniere2007}, where optimal linear combinations of reference frames are constructed in sequence for different local regions of the image space. The other method is based on Principal Component Analysis (PCA) in recent implementations such as PynPoint \citep{amara2012} and KLIP \citep{soummer2012}, where an orthogonal basis set of PCA modes is constructed and a projection of the target frame on the set is used for constructing the reference. In \citet{janson2012}, we used a loss-free implementation of LOCI. In this study, we use primarily a PCA implementation based on the KLIP procedure. We have analyzed the first epoch Fomalhaut data set of \citet{janson2012} with the PCA-based method in addition to the original LOCI implementation, and there is no significant difference in the performance of the two, but the PCA method runs faster, hence that is what is used here. For comparison and for evaluation of, for instance, possible point sources near the significance threshold of 5$\sigma$, we also perform an independent PSF subtraction of each target using the PynPoint subtraction scheme. While the two are qualitatively similar, the PynPoint reduction has a worse signal-to-noise ratio by approximately a factor 2, and we therefore focus primarily on the KLIP-based analysis here. All quantified values are based on the KLIP analysis if not stated otherwise.

As a basic criterion for a given library frame to be selected as a reference frame to the target frame, we set the condition that a hypothetical companion at 6\arcsec\ separation must have moved by at least 1~FWHM from the differential rotation, to avoid self-subtraction. In order to avoid saturated regions, we exclude the regions inside a 20 pixel radius in the case of Fomalhaut and $\epsilon$~Eri, and 40 pixels in the case of Vega, and since we are primarily interested in the PSF noise-limited regime relatively close to the star where we can substantially improve on previous work, we also limit the optimization to within 130 pixels from the star center for Fomalhaut and $\epsilon$~Eri, and within 150 pixels from the star center for Vega. In order to keep array sizes manageable, we make separate PCA subtractions for different regions: In the case of Fomalhaut and $\epsilon$~Eri, we use three regions with the first between 20 and 50 pixels in radius, the second between 50 and 100 pixels, and the third between 100 and 130 pixels. For Vega, we use two regions where the first extends from 40 to 100 pixels and the second extends from 100 to 150 pixels. The boundaries of the full concatenated optimization regions correspond to 19--125~AU for $\epsilon$~Eri, 46--300~AU for Fomalhaut, and 92--347~AU for Vega. Each PSF subtraction uses the first 100 PCA modes by default. The reduction results are not strongly dependent on the number of modes chosen.  

After subtraction, all the individual images are de-rotated so that North points in the positive y-direction and East in the negative x-direction. A median combination was used to generate one final frame per target. In order to get a broad general view of the full field of view outside of the optimization regions, we also produce ADI-subtracted full-field frames where only the mean PSF is used as a PSF reference (again after requiring at least 1~FWHM of motion between target and reference frames), with no LOCI or PCA-based optimization. Since all of the full frame fields have several point sources in them, the 3.6~$\mu$m and 4.5~$\mu$m frames are carefully compared to check if any very red objects exist among them. In order to reproduce the total infrared flux of the point sources in question, any real physical companion would have to be more than an order of magnitude fainter in the  3.6~$\mu$m band than the 4.5~$\mu$m band \citep[e.g.][]{spiegel2012}. With the exception of a few point sources around Vega that will be discussed in Sect. \ref{s:vega}, all significant point sources could be recovered at both wavelengths and none were more than a factor of $\sim$2 fainter in the 3.6~$\mu$m band than the 4.5~$\mu$m band, hence these point sources are all probable background stars. While there are no known contaminant point source objects that are as red as $<$400~K companions in these bands, we nonetheless consider it necessary for any candidate to be tested for common proper motion before anything can be said with confidence about its nature.

An actual companion in the data would inevitably suffer some partial flux loss during the PSF subtraction, so it is vital to robustly estimate the actual throughput of such a companion in order to be able to evaluate the real contrast performance of the algorithm. We use the procedure suggested as part of KLIP in \citet{soummer2012} where an image of an artificial companion is projected on the basis set that was constructed from the reference images. For Fomalhaut and $\epsilon$~Eri, the artificial companion is sequentially placed at every combination of eight different position angles (0$^{\rm o}$, 45$^{\rm o}$, 90$^{\rm o}$ etc.) and seven separations (five at 25--45 pixels in steps of 5 pixels, one at 75 pixels and one at 115 pixels). We then calculate the throughput (flux measured in a 1 FWHM diameter circular aperture centered on the companion before versus after subtraction) for all cases and evaluate the mean and median of the azimuthal points for each radial step, in order to get composite throughput estimations as a function of separation. Since there is quite a lot of azimuthal structure in the images, with the six spider arms and other asymmetric PSF features, the mean and the median can give quite different results, with the median predicting a higher throughput at most separations (e.g. 95.6\% versus 85.1\% at a 45 pixel separation), except at the smallest radii where the opposite is sometimes true (e.g. the mean predicting 48.2\% and the median predicting 47.0\% at a 25 pixel separation). In the course of this study we will use the mean estimator throughout. For Vega, the procedure is exactly the same as for the other two targets, except that the separations at which the throughput is evaluated are at 50--90 pixels in steps of 10 pixels, and at 125 pixels.

In order to evaluate the sensitivity in the final images, we calculate the standard deviation of pixels in concentric annuli around the star with widths of 1 pixel, from the inner to the outer edge of the optimization region. These are then related to the zero-point flux of Vega to acquire sensitivites in Vega magnitudes \citep[see][]{marengo2009}. We use 5$\sigma$ detection limits throughout this analysis. The throughput of a signal at each pixel separation is calculated through a linear interpolation between the points at which it is explicitly evaluated (as described above) for the innermost range (20--50 pixels for Fomalhaut and $\epsilon$~Eri, 40--100 pixels for Vega). For the 50--100 and 100--125 pixel ranges (100--150 pixels for Vega), the throughput at the center separation in the range is adopted. These throughputs are included in the sensitivity estimates. All curves presented in subsequent sections of this paper correspond to the 4.5~$\mu$m data, since this provides by far the best sensitivity to any planetary companions that may reside in the systems. Mass and effective temperature sensitivities are estimated from the flux sensitivities using evolutionary COND-based models \citep{allard2001,baraffe2003}\footnote{Photometric model values in the IRAC bands are available online via \url{http://phoenix.ens-lyon.fr/Grids/AMES-Cond/}} and the age limits for the respective targets discussed in the following section. It can be noted that the classical discrepancy between so-called hot- and cold-start models \citep[e.g.][]{marley2007,spiegel2012} has no relevance for any companions that could be detected in this study, since they would be far too old and cold to retain any memory of their initial entropy. Aside from our choice of a 5$\sigma$ significance criterion instead of a 3$\sigma$ one, the only difference between how the sensitivity is evaluated here versus how it is evaluated in \citet{marengo2009} is that we are evaluating the standard deviation in single-pixel annuli, whilst \citet{marengo2009} evaluate it in annuli of 2~FWHM widths. These methods are equivalent for the purpose of determining a characteristic contrast at a given separation. The only difference is that a wider annulus gives a coarser spatial sampling on one hand, and less point-to-point scatter in the sensitivity curve on the other hand.

\section{Results and Discussion}
\label{s:results}

In this section, we discuss separately the results acquired for the three targets observed in the study.

\subsection{$\epsilon$ Eridani}
\label{s:eri}

\begin{figure}[htb]
\centering
\includegraphics[width=8cm]{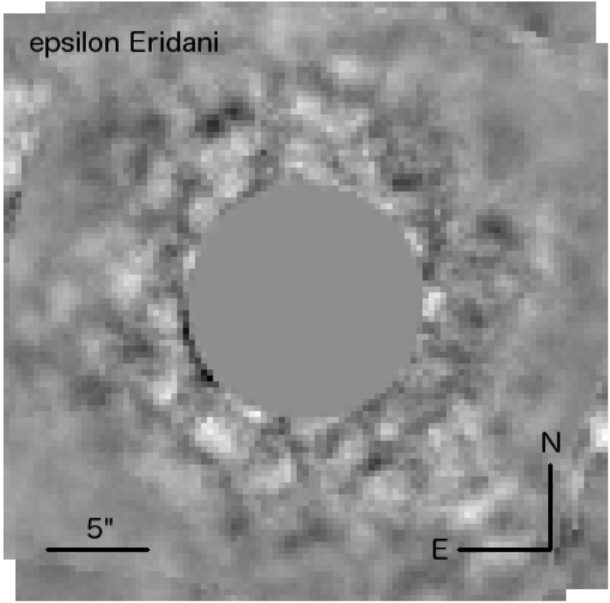}
\caption{Final reduced 4.5~$\mu$m image of the innermost region of the $\epsilon$~Eri system. No significant planetary candidates are present.}
\label{f:imeri}
\end{figure}

\begin{figure*}[htb]
\centering
\includegraphics[width=16cm]{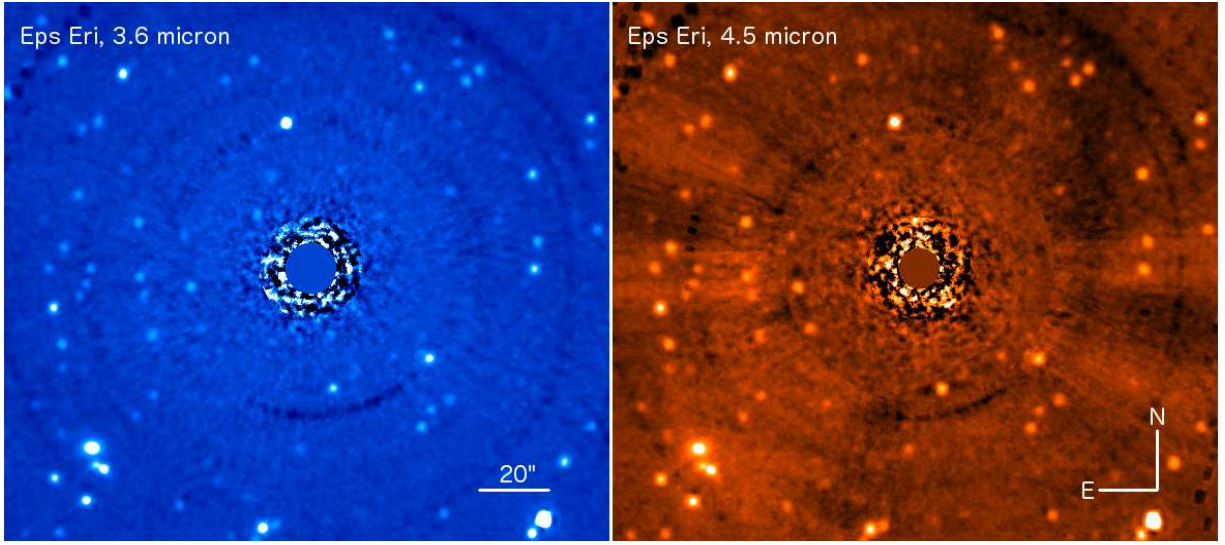}
\caption{Final reduced composite image of a wider field around $\epsilon$~Eri, at both 3.6~$\mu$m (left) and 4.5~$\mu$m (right). All significant point sources have detectable flux in both channels, while planets in the system would only be detectable in the 4.5~$\mu$m channel. In the outer ranges of the image where PCA-based subtraction has not been applied, the imperfect background removal from a simple median subtraction leads to an asymmetric background distribution between the upper right and lower left sections of the image.}
\label{f:rberi}
\end{figure*}

\begin{figure}[htb]
\centering
\includegraphics[width=8cm]{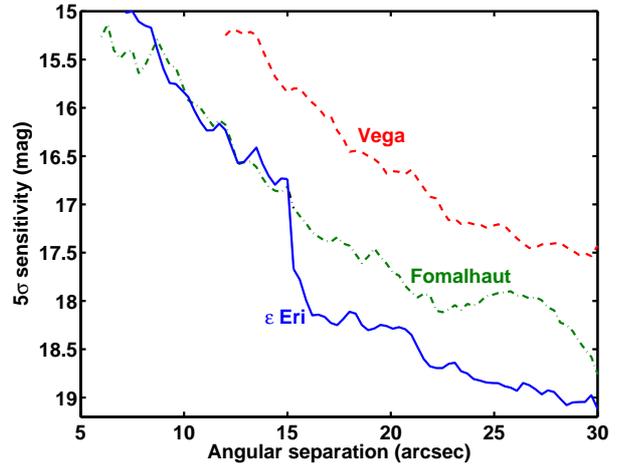}
\caption{5$\sigma$ sensitivity limits in terms of apparent magnitude in the 4.5~$\mu$m band, for the targets $\epsilon$~Eri (blue, solid line), Fomalhaut (green, dash-dotted line), and Vega (red, dashed line). Since Vega is brighter than the two other targets, it is less sensitive to very faint companions due to the bright PSF wings. Fomalhaut is slightly brighter than $\epsilon$~Eri, but since a larger amount of data is accessible for the case of Fomalhaut (see Sect. \ref{s:fomal}), the contrast performance is increased, and so for the inner parts of the separation range an approximately equal performance is attained for those two targets.}
\label{f:fluxlimits}
\end{figure}

No significant very red candidates were discovered in the final PCA-reduced image of $\epsilon$ Eridani (see Figures \ref{f:imeri} and \ref{f:rberi}), but the achieved sensitivity was excellent, allowing to place strong upper limits on the flux (and thus temperature and mass) of any wide companions in the system. The sensitivity curve is plotted in Fig. \ref{f:fluxlimits}. This sensitivity can be compared to the previous detection limits in the \citet{marengo2009} study, by noting that the 3$\sigma$ sensitivity is given as 13.60~mag at 10\arcsec\ and 14.55~mag at 15\arcsec\ there. Translating this into 5$\sigma$ values results in 13.05~mag at 10\arcsec\ and 14.00~mag at 15\arcsec. By comparison, the corresponding sensitivities acquired here are 15.89~mag and 16.73~mag, respectively. This is more than an order of magnitude increase in contrast, very similar to what we demonstrated in \citet{janson2012} for Fomalhaut. Hence, this demonstrates once again that the use of multi-angle ADI and a large number of reference frames can significantly enhance the contrast performance in \textit{Spitzer} data relative to standard two-angle differencing. 

\begin{figure}[htb]
\centering
\includegraphics[width=8cm]{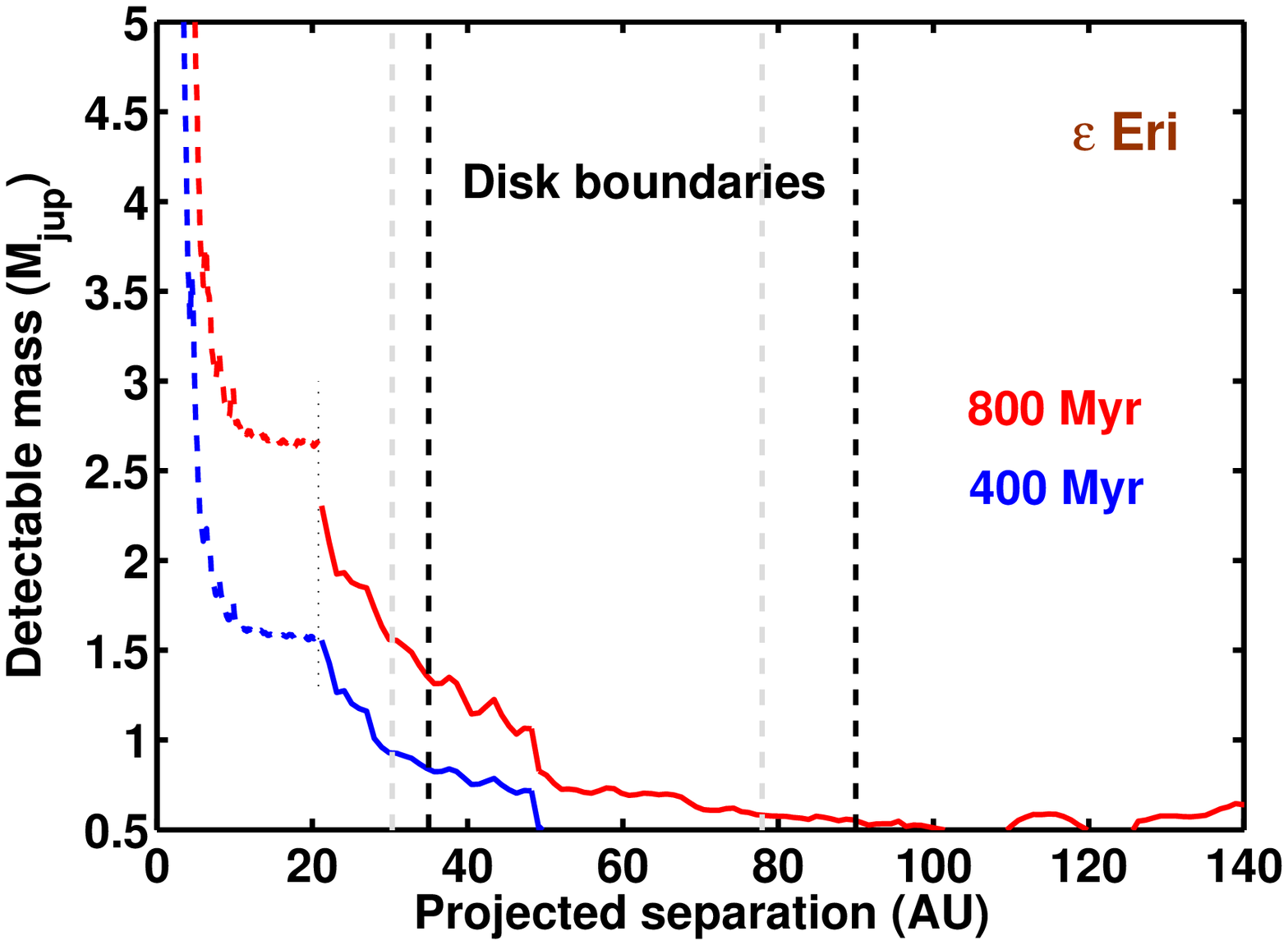}
\caption{5$\sigma$ sensitivity limits in terms of detectable planet mass for $\epsilon$~Eri, based on COND models. The models only extend down to 0.5~$M_{\rm jup}$, thus the graph cuts off at that point. Aside from the \textit{Spitzer} high-contrast data represented here with solid lines, the plot also shows, as dashed lines, the sensitivity from \citet{janson2008} at smaller separations. The lower, blue points represent an age of 400~Myr, and the upper red lines an age of 800~Myr. The transition point between the different data sets is marked with a thin dashed line. Also plotted with vertical dashed lines are the inner and outer edges of the wide debris ring in the system. Gray lines are the minimum projected separations for a system inclination of $\sim$30$^{\rm o}$ \citep{saar1997,greaves1998}. The full field stretches out to 400~AU with a roughly uniform sensitivity-limited performance, but has been cut off here to highlight the inner regions. When these imaging limits are combined with limits from radial velocity data \citep[e.g.][]{hatzes2000,zechmeister2014}, it can be concluded that no planets more massive than 3~$M_{\rm jup}$ can exist anywhere inside of $\sim$500~AU in the system.}
\label{f:masseri}
\end{figure}

Adopting an age range for $\epsilon$~Eri is necessary for  formulating the sensitivity in terms of detectable companion mass. In \citet{janson2008}, an examination of several different age diagnostics from the literature resulted in an age range of 200--800~Myr. However, the lower end of this range comes from a purely kinematic analysis \citep{fuhrmann2004}, which compared to more recent and in-depth studies related to young co-moving associations \citep[e.g.][]{schlieder2012,malo2013} must be considered as a rather loose constraint. Hence, we disregard it for the purpose of this study. In the meantime, \citet{mamajek2008} performed a study of activity and gyrochronology in nearby stars including $\epsilon$~Eri, deriving age estimates of 400~Myr and 800~Myr with two different methods. These estimates are in excellent agreement with the remaining range of ages considered in \citet{janson2008}, and so for this study, we simply adopt an age range of 400--800~Myr. Given this age range, we derive model-dependent mass detection limits as shown in Fig. \ref{f:masseri}. 

In \citet{janson2008}, it was shown from a combination of imaging and radial velocity constraints that no planets more massive than 3~$M_{\rm jup}$ could reside anywhere in the system, at least inside of the $\sim$500~AU field radius of \textit{Spitzer}, even at the upper bound of the system age. Our new Spitzer limit further underlines this conclusion, and strengthens it with yet tighter constraints at wide separations. Even sub-Jovian planets can be discovered in some parts of the system -- from $\sim$28~AU and outwards for young system ages and from $\sim$48~AU and outwards for old ages. As mentioned previously, it is of particular interest to study the regions close to the inner and outer edges of the debris ring. In the case of $\epsilon$~Eri, the ring extends from approximately 35~AU to 90~AU \citep{backman2009}. Thus, including projection effects, $\sim$30--35~AU is the interesting range for the case of the inner edge. Here, we can exclude planets more massive than 0.8--0.9~$M_{\rm jup}$ if the system age is 400~Myr, and planets more massive than 1.3--1.6~$M_{\rm jup}$ if the age is 800~Myr. For the case of the outer edge, we can exclude planets more massive than 0.6~$M_{\rm jup}$ at 78--90~AU if the age is 800~Myr. At a 400~Myr age, we are sensitive to planets less massive than 0.5~$M_{\rm jup}$ in this separation range, but since no models exist below 0.5~$M_{\rm jup}$, a more specific value cannot be assigned at this point. 

\begin{figure}[htb]
\centering
\includegraphics[width=8cm]{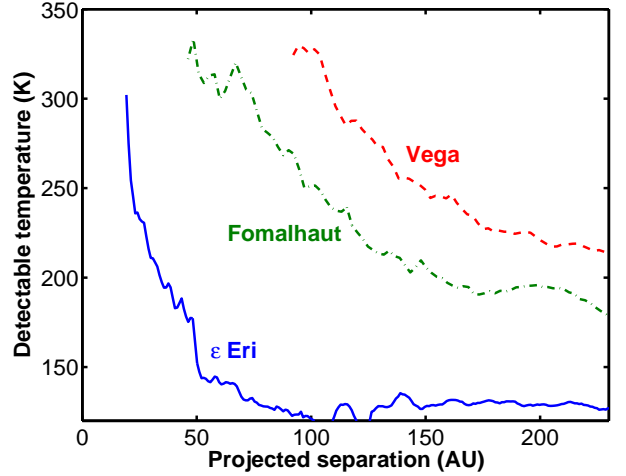}
\caption{5$\sigma$ sensitivity limits in terms of detectable effective temperature of a Jupiter-sized planet based on COND models, for the targets $\epsilon$~Eri (blue, solid line), Fomalhaut (green, dash-dotted line), and Vega (red, dashed line). Because $\epsilon$~Eri is the closest of the three targets at 3.3~pc, it offers sensitivity to the lowest temperatures. Planets even as cold as Jupiter (134~K) can be detected at $>$80~AU in the system.}
\label{f:templimits}
\end{figure}

The sensitivities in terms of effective temperatures for the three targets are shown in Fig. \ref{f:templimits}. Since $\epsilon$~Eri is the most nearby of the targets, it provides the best temperature sensitivity. Near the inner edge of the debris ring, $\sim$200~K planets are detectable, and the background-limited sensitivity from about 80~AU and outwards is 120--130~K. This is even lower than the effective temperature of Jupiter \citep[134~K, see][]{aumann1969}. Hence, planets with masses and atmospheres identical to Jupiter (aside from factors related to insolation) are directly detectable with \textit{Spitzer} at separations of 80~AU and beyond in the $\epsilon$~Eri system. This is a thoroughly unique feature of this target (and telescope), as typically only significantly hotter planets can be detected with state-of-the-art facilities. For reference, the coldest exoplanet imaged so far, GJ~504~b \citep{kuzuhara2013}, has an effective temperature in the range of 500--600~K \citep{janson2013b}.

\subsection{Fomalhaut}
\label{s:fomal}

\begin{figure*}[htb]
\centering
\includegraphics[width=16cm]{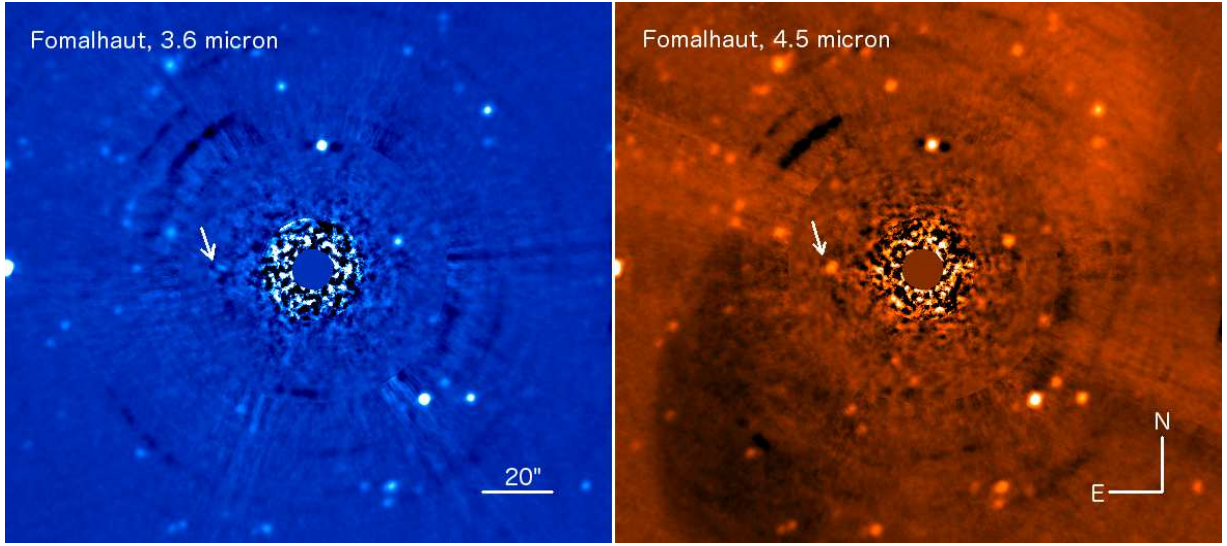}
\caption{Final reduced second epoch composite image of a wider field around Fomalhaut, at both 3.6~$\mu$m (left) and 4.5~$\mu$m (right). As for the case of $\epsilon$~Eri, all point sources in the image have detectable signatures in both channels, consistent with background stars. The point source marked with an arrow has a strange morphology at 3.6~$\mu$m, possibly due to it coinciding with a residual spider feature. In the outer ranges of the image where PCA-based subtraction has not been applied, the imperfect background removal from a simple median subtraction leads to an asymmetric background distribution between the upper right and lower left sections of the image.}
\label{f:rbfomal}
\end{figure*}

\begin{figure}[htb]
\centering
\includegraphics[width=8cm]{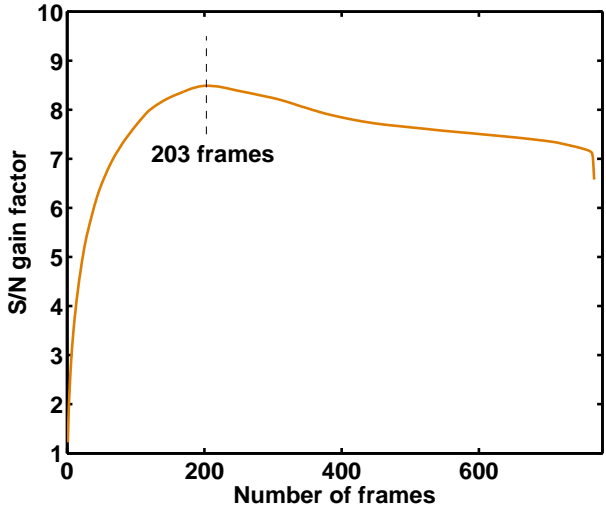}
\caption{$S/N$ gain factor as a function of cumulative addition of individually PSF-subtracted frames sorted from least to greatest individual scatter. As the frames become progressively worse, they add less and less to the final $S/N$, and eventually decrease the quality rather than increasing it. It follows that the best 203 frames provide the ideal ensemble to combine for optimizing the total $S/N$.}
\label{f:snfactor}
\end{figure}

For Fomalhaut, the quality of the PCA-reduced data of the second epoch alone (see Fig. \ref{f:rbfomal}) is significantly worse than for the first epoch presented in \citet{janson2012}. The cause of this is unclear. A PCA re-reduction of the first epoch data confirms that this is an intrinsic feature of the data and not due to the reduction. The difference is large enough that co-adding the full two epochs offers no improvement in $S/N$ versus using the first epoch by itself. Examining the individual frames after PCA reduction but before median collapse shows that indeed, the individual standard deviations of pixels in the optimization region are higher on average for the second epoch than the first epoch. However, there is a significant spread in the scatter among individual frames in both epochs, such that some second epoch frames still exhibit smaller scatter than some first epoch frames. \footnote{The spread in scatter is a natural consequence of the PCA reduction (and other optimized reference techniques), since there is variation in how well a given PSF reference matches the target frame based on how well the particular features of the target frame matches those represented in the set of references.} As a consequence, we attempt to maximize the $S/N$ of the combination of both data sets by selecting the optimal combination of frames for this purpose. We do this by sorting the combined set of PCA-reduced frames by their scatter, and calculating the $S/N$ that would result from combining an incremental number of frames, starting from the smallest scatter and working upward. The calculation is based upon the assumption of the noise being independent between frames, such that it combines in a root-mean-square fashion. A resulting plot is shown in Fig. \ref{f:snfactor}, which demonstrates that there is indeed an optimal number of frames to be combined, after which adding more frames with incrementally larger scatter will only decrease the final $S/N$. We thus produce a median-combined frame of the 203 frames selected in this manner. As expected, the final frame does constitute a modest improvement over the first epoch alone or the full combination of both epochs, so we use this frame for all future purposes in this paper. 

\begin{figure*}[htb]
\centering
\includegraphics[width=16cm]{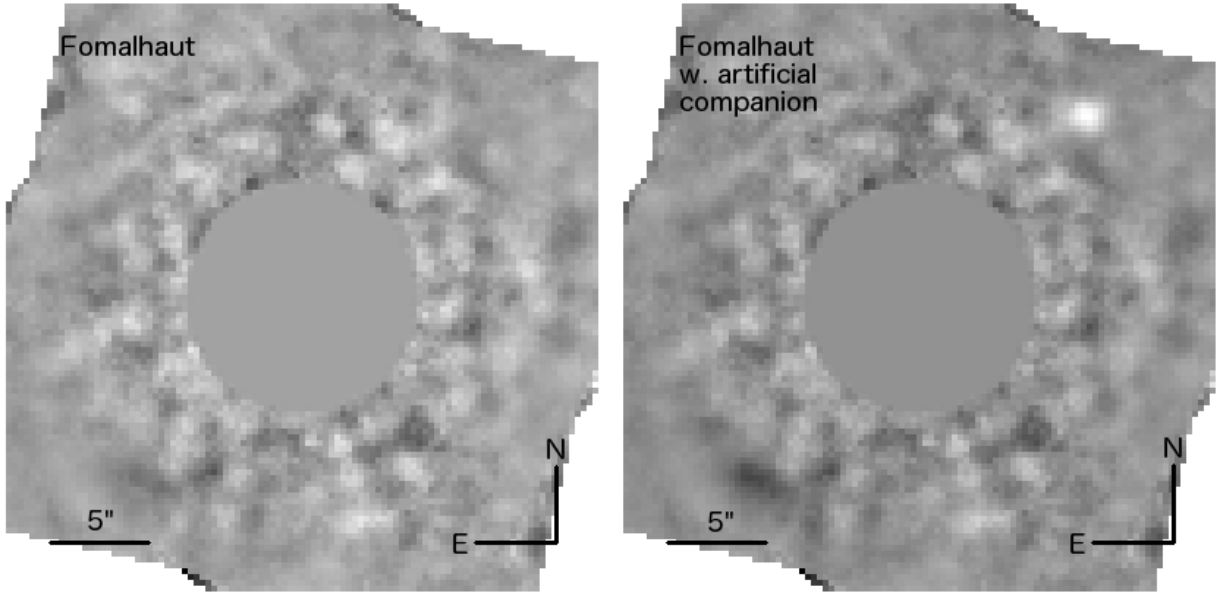}
\caption{Final reduced 4.5~$\mu$m images of the the innermost region in the Fomalhaut system. Left: The actual image, with no candidates present. Right: An image in which an artificial companion has been inserted at the expected location of Fomalhaut~b prior to the high-contrast reduction, illustrating that companions are well preserved in the procedure, with little flux loss. The artificial companion in this example corresponds to a planet of mass 1.6--1.8~$M_{\rm jup}$ at 400--500~Myr using COND models.}
\label{f:imfomal}
\end{figure*}

The final images are shown in Fig. \ref{f:rbfomal} and Fig. \ref{f:imfomal}. There are no very red candidates in the images, although the closest candidate straight to the East in Fig. \ref{f:rbfomal} has a strange appearance in the 3.6~$\mu$m channel, in that a dark streak appears to coincide with the point source. In \citet{janson2012}, we pointed out a possible point source to the South that was statistically insignificant but nonetheless worthy of some further attention. With the addition of new data, the point source disappears, confirming that it was most likely a spurious speckle. The azimuthally averaged detection limit is shown along with the other targets in Fig. \ref{f:fluxlimits}. Since it is particularly interesting to evaluate the sensitivity at the location of Fomalhaut~b \citep{kalas2008}, and since the sensitivity varies a lot across the field as noted earlier, we calculate explicitly the sensitivity at that location, through the standard deviation in a 5-by-5 pixel box. We also evaluate the throughput at that particular location by imposing a false companion in the non-reduced data and transmitting it through the reduction procedure and comparing the flux before and after. Here, the throughput is 87.9\%, which is intermediate between the median and mean throughputs at this separation. The resulting 5$\sigma$ sensitivity at the location of Fomalhaut~b is 17.3~mag, a 0.4~mag improvement on the original epoch. In \citet{janson2012}, we used the \citet{spiegel2012} models to derive a mass limit of 1~$M_{\rm jup}$ from the previous detection limit at an age of 400~Myr. The difference in the IRAC 4.5~$\mu$m band at this age between a 0.5~$M_{\rm jup}$ and a 1~$M_{\rm jup}$ planet in the \citet{spiegel2012} models is 0.8~mag, so an interpolation would imply a sub-Jovian mass in the range of 0.75~$M_{\rm jup}$ in this circumstance. The COND models predict a broadly consistent though slightly higher mass of 1.0~$M_{\rm jup}$ for an age of 400~Myr (and 1.2~$M_{\rm jup}$ for 500~Myr).

Since the full combination of two data sets spans a rather long time baseline, it is relevant to consider whether orbital motion could affect the detectability. E.g., if Fomalhaut b would have moved enough between the two epochs, its signatures from each respective epoch would not co-add, but rather there would be two separate signatures in the final image. However, this effect can be easily estimated since it is known from Hubble Space Telescope (HST) observations how the object moves. Between the two latest reported epochs from \citet{kalas2013}, the object moves at a rate of 124~mas/yr. The difference between the mean observational epoch of our two data sets is 2.4 years, leading to a total estimated motion of approximately 300~mas. This is equal to one oversampled \textit{Spitzer} pixel, and much smaller (by close to a factor 6) than the FWHM. Thus, the signatures will co-add efficiently and orbital motion of Fomalhaut~b can be considered negligible in this context.

\begin{figure}[htb]
\centering
\includegraphics[width=8cm]{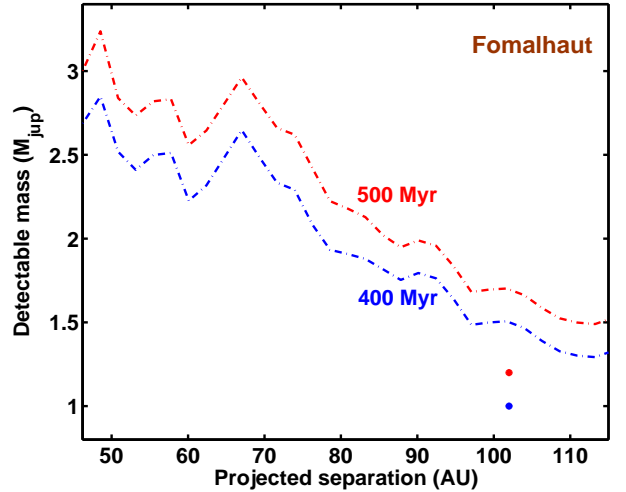}
\caption{5$\sigma$ sensitivity limits in terms of detectable planet mass for Fomalhaut, based on COND models. The lower blue (400~Myr age) and upper red (500~Myr) curves are the azimuthally averaged sensitivities. These sensitivities vary significantly over the field of view, and in the expected location of Fomalhaut~b, they are better than average. They are illustrated by points at a bit over 100~AU in the figure.}
\label{f:massfomal}
\end{figure}

The average COND-based mass detection limit is shown in Fig. \ref{f:massfomal}. Limits of 400~Myr and 500~Myr are used to bracket the age, which corresponds broadly to the age estimate of 440$\pm$40~Myr for the Fomalhaut system provided in the detailed study by \citet{mamajek2012}. The temperature detection limit is shown in Fig. \ref{f:templimits}. As can be seen, planets in the range of 1.5--3~$M_{\rm jup}$ and 250--330~K are detectable on average in the range of 50--100 AU. Although alternative scenarios have been proposed \citep[e.g.][]{lyra2013}, the morphology of the Fomalhaut disk has been proposed to possibly imply the presence of massive planets \citep[e.g.][]{quillen2006}. Thus, our study provides the most robust available upper limits for the masses of any such planets in wide ($>$50~AU) orbits. See \citet{kenworthy2013} and \citet{currie2013} for summaries of constraints from a range of observations.

As we showed already in \citet{janson2012}, the upper limit on 4.5~$\mu$m flux at the location of Fomalhaut~b firmly excludes the possibility that any noticeable fraction of the visible-light flux observed in the Hubble images constitutes thermal radiation from a giant planetary surface. Rather, the flux represents reflected emission from the star against dust in some configuration (which may in turn be associated with a planet or planet-like parent body). The upper limit is further tightened by these new observations, and in the meantime, several studies have strengthened the reflected light hypothesis by recovering the point source at yet shorter wavelengths than the original detection \citep{currie2012,galicher2013,kalas2013}. Several studies have been performed for placing constraints on the physical characteristics and origin of Fomalhaut~b \citep[e.g.][]{beust2014,kenyon2014,tamayo2014}, but nonetheless, further observational input would be highly useful to better understand this intriguing object. Whatever its precise nature, it is clear that it is an object with (as of yet) no known close counterparts in other planetary systems, and so its interest for continued future study is obvious.

\subsection{Vega}
\label{s:vega}

\begin{figure}[htb]
\centering
\includegraphics[width=8cm]{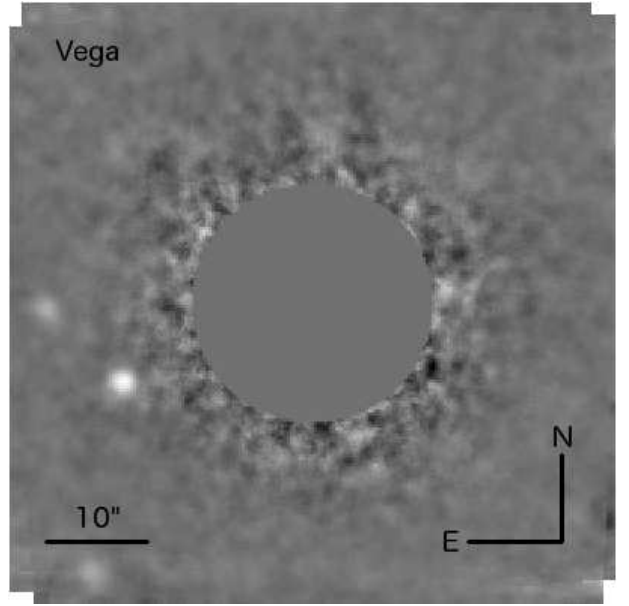}
\caption{Final reduced 4.5~$\mu$m image of the innermost region in the Vega system. The three point sources in the South-Eastern quadrant of the image are all probable background sources (see text).}
\label{f:imvega}
\end{figure}

\begin{figure*}[htb]
\centering
\includegraphics[width=16cm]{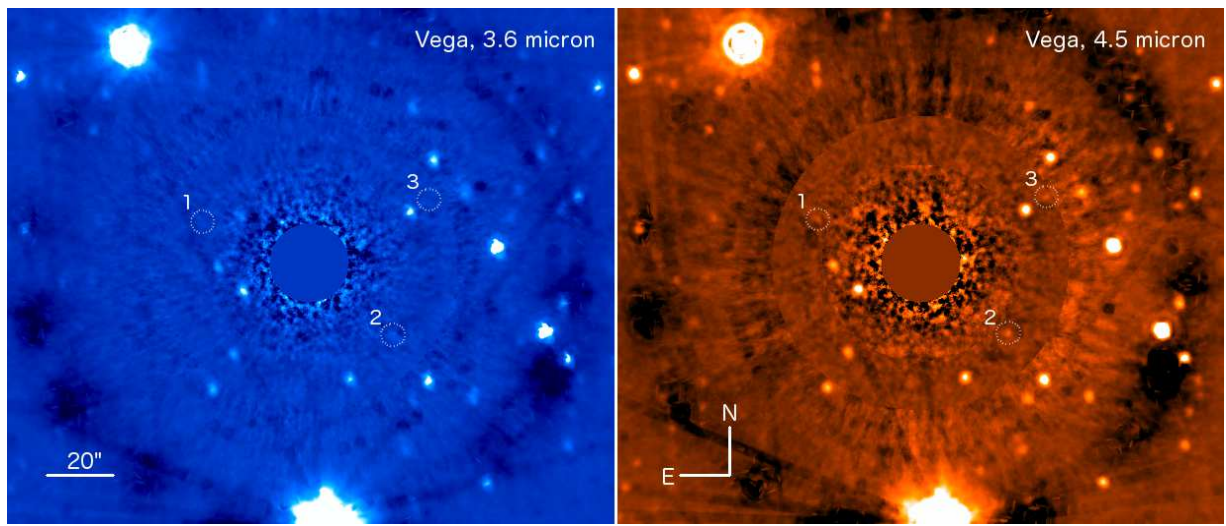}
\caption{Final reduced composite image of a wider field around Vega, at both 3.6~$\mu$m (left) and 4.5~$\mu$m (right). There are some  4.5~$\mu$m point sources for which it is not clear if a 3.6~$\mu$m counterpart is present; see discussions in the text.}
\label{f:rbvega}
\end{figure*}

Since Vega is a bit brighter than the other two targets and the observing parameters were identical, it saturates out to a farther angular separation of about 40~pixels. We thus perform the PCA-based reduction in an area of 40--150~pixel radius from the star, as described in Sect. \ref{s:psf}. The innermost region of the system is shown in Fig. \ref{f:imvega} and a wider field is shown in Fig. \ref{f:rbvega}. There are no unambiguously interesting candidates in the final reduced frame, but there are a few cases that deserve special attention. These are marked out in the latter figure. 

Feature `1' is not statistically significant (only 2.7$\sigma$) in the KLIP reduction, but it appears at 6.2$\sigma$ in the PynPoint reduction. The feature appears to reside in an angular range that is particularly affected by spider features, and so we consider it a likely spurious feature, since it is not significant in the quantitatively better KLIP reduction. For feature `2', it is unclear whether or not the 4.5~$\mu$m feature (which is significant at 5.2$\sigma$) has a 3.6~$\mu$m counterpart. Our preliminary assessment is that it is a slightly red background object, but further observations would be useful to verify this. Feature `3'  is another case of a 4.5~$\mu$m feature (5.6$\sigma$ significance) with no 3.6~$\mu$m counterpart, and is perhaps the most promising of the candidates. However, while the feature is point-like in the KLIP reduction, it appears more extended in the radial direction in the PynPoint reduction, again raising the possibility that it might be a residual spider feature. Follow-up observations would be useful for further testing all of these cases. If interpreted as real, physically bound companions, their projected separations correspond to $\sim$265--335~AU, and their fluxes correspond to masses of $\sim$2--3~$M_{\rm jup}$ at ages of 400--800Myr.

\begin{figure}[htb]
\centering
\includegraphics[width=8cm]{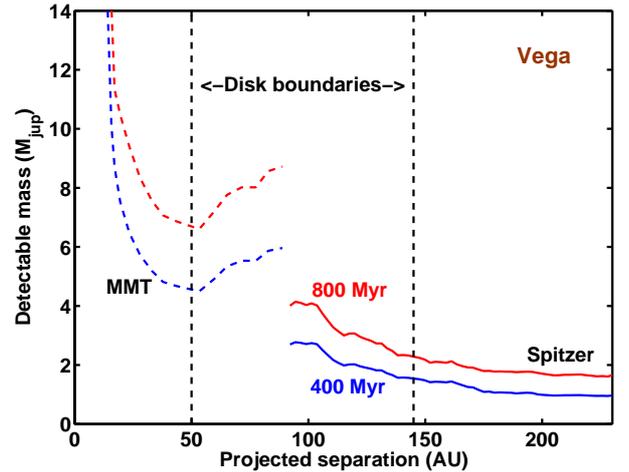}
\caption{5$\sigma$ sensitivity limits in terms of detectable planet mass for Vega, based on COND models. Aside from the Spitzer data analyzed here and plotted with solid lines, the figure also includes a limit from the MMT in dashed lines probing smaller separations in the system, by \citet{heinze2008}. Also shown with vertical dashed lines are the approximate inner and outer edges of the wide debris ring in the system. The lower blue lines correspond to a system age of 400~Myr, and the upper red lines to an age of 800~Myr.}
\label{f:massvega}
\end{figure}

The sensitivity curve for Vega is included in Fig. \ref{f:fluxlimits}. The higher limits around Vega than the other targets stem from the fact that an approximately equal contrast is achieved as for the other targets, but around a brighter target star. The age of Vega is typically estimated through isochronal techniques in the literature, and recent estimates have varied fairly significantly in the literature \citep{song2001,yoon2010,monnier2012}. Here we adopt an age range of 400--800~Myr to encompass this uncertainty. This leads to model-based mass detection limits as shown in Fig. \ref{f:massvega}, with a sensitivity to 3--4~$M_{\rm jup}$ planets at $\sim$100~AU. The disk boundaries plotted in Fig. \ref{f:massvega} are based on the estimated half-maximum points of the 1~mm optical depth in \citet{marsh2006}. As argued in e.g. \citet{boley2012}, the relatively large grains probed at such wavelengths are probably more closely representative of the parent planetesimal body distribution than dust probed at shorter wavelengths \citep[e.g.][]{sibthorpe2010}, which imply a somewhat different dust distribution. When combining the Spitzer data to literature M-band limits \citep{heinze2008} using the same ages and the same evolutionary and atmospheric models, the mass limits close to the disk boundaries are 5--7~$M_{\rm jup}$ at the inner edge at $\sim$50~AU, and $\sim$2~$M_{\rm jup}$ at the outer edge near 145~AU. The detection limit in terms of effective temperature is included in Fig. \ref{f:templimits}. While not quite as sensitive in this regard as the Fomalhaut and $\epsilon$~Eri limits, the Vega observation still allows for detection of objects with substantially lower temperatures than can be acquired at shorter wavelengths with currently existing instrumentation.

% Figures: 1) Sensitivity versus angular separation (common to the three targets); 2) high-contrast image of Vega; 3) Mass sensitivity versus proj. physical separation along with the disk boundaries and the Heinze limits; 4) temperature versus projected physical separation (common for the three stars).

\section{Summary and Conclusions}
\label{s:summary}

In this study, three stars that are both nearby, young, and have large debris disks with inner gaps were studied with \textit{Spitzer} using high-contrast methods, in order to yield unprecedented sensitivity to cold, low-mass companions at wide separations. One of these targets was Fomalhaut, which had already been studied with the same technique in \citet{janson2012} and which motivated this study. Although our new Fomalhaut data set was of worse quality on average than the previous epoch of data, a combination of the two data sets with selection of the best frames yielded improved sensitivity limits. In terms of mass sensitivity, the limits did not change much, as the improved flux sensitivity was counteracted by the fact that the system is now thought to be a bit older than previously believed \citep{mamajek2012}. Thermal radiation from Jupiter-mass or more massive planets can be excluded at the expected location of Fomalhaut~b. Another observed target was Vega, for which we were sensitive to planets more massive than $\sim$2~$M_{\rm jup}$ outside of the outer disk edge at 145~AU. A few candidates detected in the 4.5~$\mu$m image would greatly benefit from follow-up observations in the future aiming for a similar contrast performance, in order to better establish their nature. Particularly tight constraints on planetary properties could be set in the $\epsilon$~Eri system, due to its proximity. Prior to the observations presented here, it was possible to exclude that any planets more massive than 3~$M_{\rm jup}$ exist anywhere in the system inside of $\sim$500~AU \citep{janson2008}, based on a combination of radial velocity and imaging data. With these new \textit{Spitzer} data, it was possible to place even stronger constraints at wide separations: For instance, planets more massive than 1.5~$M_{\rm jup}$ could be excluded at the inner edge of the $\epsilon$~Eri debris ring at 30--35~AU, and substantially sub-jovian mass planets could be excluded beyond the outer edge at 78--90~AU. Generally, planets with both the same mass (and thus equal surface gravity, for an equal radius) and same effective temperature as Jupiter could be excluded at wide separation, which is a thoroughly unique feature for this target.

These \textit{Spitzer} observations probe a new parameter range of cold and wide planets, which are unattainable with any other existing telescope or instrument. In this way, it paves the way for the James Webb Space Telescope, which will offer observations in the same wavelength range in space but with a 6~m aperture instead of the 0.85~m aperture of \textit{Spitzer}, greatly improving the sensitivity and spatial resolution. More generally, the results attained in \citet{janson2012} and here, as well as the similar results from HST data \citep[e.g.][]{lafreniere2009,soummer2014}, demonstrate the great benefit of applying high-contrast techniques and algorithms to space-based telescopes, with their high degree of PSF stability, which enables sophisticated PSF reference optimization. This has broad utility for high-contrast imaging, potentially including more advanced coronagraphy/occulter-based missions for imaging Earth-like planets in the habitable zones of nearby stars at some point in the future.

\begin{acknowledgements}
M.J. gratefully acknowledges funding from the Knut and Alice Wallenberg Foundation. J.C. receives support from the Research Corporation for Science Advancement (Award No. 21026) and the South Carolina Space Grant Consortium. This work is based on observations made with the Spitzer Space Telescope, which is operated by the Jet Propulsion Laboratory, California Institute of Technology under a contract with NASA. This study made use of the CDS services SIMBAD and VizieR, as well as the SAO/NASA ADS service.
\end{acknowledgements}

\end{document}